[1994/12/01]

\documentclass[colordvi]{amsart}

\newif\ifpdf\ifx\pdfoutput\undefined\pdffalse\else\pdfoutput=1\pdftrue\fi

\usepackage{amscd}
\usepackage{amsthm}   
\usepackage{amsfonts}
\usepackage{amssymb}
\usepackage{amsmath}
\usepackage{verbatim}
\usepackage{alltt}    
\usepackage{color}

\ifpdf
\usepackage[pdftex,colorlinks,citecolor=blue]{hyperref} 
\else
\usepackage[hypertex,citecolor=blue]{hyperref} 
\fi

\newcommand{\ket}[1]{\ensuremath{|#1\rangle}}

\newcommand{\braket}[2]{\ensuremath{\langle #1 |#2\rangle}}

\newcommand{\TODO}[1]{} 

\newtheorem{thm}{Theorem}
\newtheorem{theorem}[thm]{Theorem}
\newtheorem{defn}[thm]{Definition}
\newtheorem{lemma}[thm]{Lemma}
\newtheorem{cor}[thm]{Corollary}

\newcommand{\round}[1]{\ensuremath{\left\lfloor #1\right\rceil}}  
\newcommand{\floor}[1]{\ensuremath{\left\lfloor #1\right\rfloor}} 
\newcommand{\ceil}[1]{\ensuremath{\left\lceil #1\right\rceil}}    

\newcommand{\Hales}{\cite{Hal02}}       
\newcommand{\Hallgren}{\cite{HalHal00}} 

\begin{document}

\title{A Quantum Fourier Transform Algorithm}
\author{Chris Lomont}
\date{March 2004}
\email{clomont@cybernet.com, clomont@math.purdue.edu}

\thanks{Research supported by AFRL grant F30602-03-C-0064}
\urladdr{www.math.purdue.edu/$\tilde{~}$clomont}\urladdr{www.cybernet.com}
\curraddr{Cybernet Systems Corporation\\
727 Airport Blvd.\\
Ann Arbor, MI, 48108-1639 USA.}



\begin{abstract}
Algorithms to compute the quantum Fourier transform over a cyclic
group are fundamental to many quantum algorithms. This paper
describes such an algorithm and gives a proof of its correctness,
tightening some claimed performance bounds given earlier. Exact
bounds are given for the number of qubits needed to achieve a
desired tolerance, allowing simulation of the algorithm.
\end{abstract}
\maketitle

\section{Introduction}
Most quantum algorithms giving an exponential speedup over
classical algorithms rely on efficiently computing Fourier
transforms over some finite group
\cite{BPR99,CEMM98,GSVV01,Kitaev95,KaufLom02,MRRS02}. The Abelian
group case depends on fast quantum algorithms for doing Fourier
transforms over cyclic groups \cite{Hal02,Mos99,MZ03}. The thesis
\Hales~and the paper \Hallgren~describe such a quantum algorithm,
but the proofs are incorrect. This note attempts to correct those
proofs, and in the process obtains stronger bounds for many of
their results, and a few weaker ones. The end result is a proof of
the correctness of their algorithm, with concrete bounds suitable
for quantum simulation instead of the asymptotic bounds listed in
their papers. Our final result is theorem \ref{t:MainBound}.

\section{Preliminaries}
Efficient algorithms for the quantum Fourier transform over finite
Abelian groups are constructed from the algorithms for the
transform over cyclic groups, which in turn reduce to computing
the transform efficiently over prime order groups
\cite{Hal02,Mos99}. Efficient algorithms for computing the quantum
Fourier transform over a cyclic group of order $2^m$ for a
positive integer $m$ are well known, and are used in Shor's
factoring algorithm: see Coppersmith \cite{Copper94} and Shor
\cite{Shor94,Shor97} for example. We will show an algorithm for
computing the quantum Fourier transform over an odd order cyclic
group. The algorithm (containing minor errors) is given in
\Hales~and \Hallgren, but their proofs of the correctness of the
algorithm and subsequent performance bounds are incorrect. This
paper corrects those proofs and obtains new bounds. For
applications of this algorithm, see \Hales~ and \Hallgren. A proof
of similar ideas using a different method is in \cite{Hoy00}.

\subsection{Notation and basic facts}
We fix three integers: an odd integer $N\geq 3$, $L\geq 2$ a power
of 2, and $M\geq LN$ a power of 2. This gives $(M,N)=1$, which we
need later.

Some notation and facts to clarify the presentation:
\begin{itemize}
\item $\sqrt{-1}$ will be written explicitly, as $i$ will always
denote an index.

\item For an integer $n>1$, let $\omega_n=e^{2\pi\sqrt{-1}/n}$
denote a primitive $n^\text{th}$ root of unity.

\item Fact:
$\left|1-e^{\theta\sqrt{-1}}\right|\leq\left|\theta\right|$ as can
be seen from arc length on the unit circle. If
$-\pi\leq\theta\leq\pi$ we also\footnote{This range can be
extended slightly.} have
$\left|\frac{\theta}{2}\right|\leq\left|1-e^{\theta\sqrt{-1}}\right|$.
Thus for real values $\alpha$ we have
$\left|1-\omega_M^\alpha\right|\leq\left|\frac{2\pi\alpha}{M}\right|$,
etc.

\item $\log n$ denotes $\log$ base 2, while $\ln n$ is the natural
log. Since $M$ and $L$ are powers of two, $\ceil{\log
M}=\floor{\log M}=\round{\log M}=\log M$, and similarly for $L$,
but we often leave the symbols to emphasize expressions are
integral.

\item For a real number $x$, \ceil{x} is the smallest integer
greater than or equal to $x$, \floor{x} is the largest integer
less than or equal to $x$, and \round{x} is the nearest integer,
with ties rounding up\footnote{We could break ties arbitrarily
with the same results.}. We often use the three relations:
\begin{align*}
x-\frac{1}{2}&\leq\round{x}\leq x+\frac{1}{2}\\
x-1&<\floor{x}\leq x\\
x&\leq\ceil{x}<x+1
\end{align*}

\item Indices: $i$ and $s$ will be indices from $0,1,\dots,N-1$.
$j$ will index from $0,1,\dots,L-1$. $k$ will index from
$0,1,\dots,M-1$. $a$ and $b$ will be arbitrary indices. $t$ will
index from a set $C_s$, defined in definition \ref{d:delta} below.

\item Given $i\in\{0,1,\dots,N-1\}$, let $i'=\round{\frac{M}{N}i}$
denote the nearest integer to $\frac{M}{N}i$ with ties broken as
above. Similarly for $s$ and $s'$. Note $0\leq i'\leq M-1$.

\item For a real number $x$ and positive real number $n$, let $x
\bmod n$ denote the real number $y$ such that $0\leq y<n$ and $y =
x+mn$ for an integer $m$. Note that we do not think of $x\bmod n$
as an equivalence class, but as a real number in $[0,n)$.

\item \ket{u} and \ket{v} are vectors in spaces defined later, and
given a vector \ket{u} denote its coefficients relative to the
standard (orthonormal) basis $\{\ket{0},\ket{1},\dots,\ket{n-1}\}$
by $u_0,u_1,\dots,u_{n-1}$, etc.

\item For a real number $x$, let
\begin{equation*}
\left|x\right|_M=\left\{\begin{matrix}x\bmod M&\text{ if }& 0\leq
(x\bmod M)\leq \frac{M}{2}\\-x\bmod M&\text{
otherwise}&\end{matrix}\right.
\end{equation*}
Thus $0\leq |x|_M\leq\frac{M}{2}$. Properties of this function are
easiest to see by noting it is a sawtooth function, with period
$M$, and height $M/2$.

\item For an integer $s$ set
$\delta_s=\round{\frac{M}{N}s}-\frac{M}{N}s$. Then
$\left|\delta_s\right|\leq\frac{1}{2}$.

\item The (unitary) Fourier transform over a cyclic group of order
$N$ is denoted $F_N$. Thus if
$\ket{u}=\sum_{i=0}^{N-1}u_i\ket{i}$, then
$F_N\ket{u}=\frac{1}{\sqrt{N}}\sum_{i,s=0}^{N-1}u_i\omega_N^{is}\ket{s}$.
We write $\ket{\hat{u}}=F_N\ket{u}$, with coefficients
$\hat{u}_i$.

\item $\sum_{i=0}^{N-1} |u_i|^2=1$ implies
$\sum_i|u_i|\leq\sqrt{N}$.
\end{itemize}
~\\

The majority of the errors in \Hales~and \Hallgren~resulted from
misunderstanding the consequences of their versions of the
following two definitions. The first defines sets of integers
which will play an important role:
\begin{defn}\label{d:intervals}
For $i=0,1,\dots,N-1$, let $(i)$ denote the set of integers in the
open interval
$\left(i'-\frac{M}{2N}+\frac{1}{2},i'+\frac{M}{2N}-\frac{1}{2}\right)$
taken $\bmod\;M$. Recall $i'=\round{\frac{M}{N}i}$.
\end{defn}

The second definition we make precise is a division and remainder
operation:
\begin{defn}\label{d:delta}
Given $M,N$ as above. Set
$\alpha=\floor{\frac{M}{2N}+\frac{1}{2}}$, and
$\beta=\ceil{\frac{M}{2N}-\frac{3}{2}}$. We define the map
$\Delta:\{0,1,\dots,M-1\}\rightarrow\{0,1,\dots,N-1\}\times\{-\alpha,-\alpha+1,\dots,\alpha\}$,
as follows: for any $k\in\{0,1,\dots,M-1\}$, let
$k\xrightarrow{\Delta}(s,t)$, via
\begin{eqnarray*}
k'&=&\round{k\frac{N}{M}}\\
t&=&k-\round{k'\frac{M}{N}}\\
s&=& k'\bmod N
\end{eqnarray*}
We extend this definition to a transform of basis elements
$\ket{k}$ via
\begin{eqnarray*}
\Delta\ket{k}=\ket{s}\ket{t+\alpha}
\end{eqnarray*}
and extend to all vectors by linearity.

Finally, from the image of $\Delta$, define $C_s=\{\;t\;\;
\big|\;\; (s,t)\in \text{\emph{Image }}\Delta\}$ to be those
values of $t$ appearing for a fixed $s$. Thus
$\sum_{k=0}^{M-1}\ket{k}\xrightarrow{\Delta}\sum_{s=0}^{N-1}\sum_{t\in
C_s}\ket{s}\ket{t+\alpha}$.
\end{defn}

We will show the integers $\{-\beta,\dots,\beta\}\subseteq
C_s\subseteq\{-\alpha,\dots,\alpha\}$ for all $s$, which is why we
defined $\beta$ with the $\Delta$ definition. $\alpha$ and $\beta$
remain fixed throughout the paper.

For the proofs to work, we need that the sets $(i)$ are disjoint
and have the same cardinality. Neither \Hales~nor \Hallgren~define
these sets $\bmod\;M$, although perhaps it is implied.
\Hales~makes a similar definition without the $\frac{1}{2}$ terms,
but the resulting sets are \emph{not} then disjoint.
\Hallgren~makes a similar definition, but uses $\frac{M}{N}i$
instead of $i'$ and drops the $\frac{1}{2}$ terms, which results
in sets of varying cardinality. To see the differences, check the
three definitions using $M=32$ and $N=5$. Both \Hales~and
\Hallgren~implicitely assume their resulting sets are disjoint and
of constant cardinality in numerous places, invalidating many
proofs. Note also that the $\bmod\;M$ condition gives
$M-1,0\in(0)$ when $M>3N$. We now show that the sets defined here
have the required properties:
\begin{lemma}\label{l:setProp}
For $i_1\neq i_2\in\{0,1,\dots,N-1\}$,
\begin{eqnarray}
\left|(i_1)\right| &=& \left|(i_2)\right|\\
(i_1)\bigcap(i_2)&=&\varnothing
\end{eqnarray}
\end{lemma}
\begin{proof}
Each set is defined using an interval of constant width, centered
at an integer, so the sets will have the same cardinality. To show
disjointness, for any integer $a$, take the rightmost bound
$R_a=\round{\frac{M}{N}a}+\frac{M}{2N}-\frac{1}{2}$ of an interval
and compare it to the leftmost bound
$L_{a+1}=\round{\frac{M}{N}(a+1)}-\frac{M}{2N}+\frac{1}{2}$ of the
next interval:
\begin{eqnarray}
L_{a+1}-R_a&=&\round{\frac{M}{N}(a+1)}-\round{\frac{M}{N}a}-\frac{M}{N}+1\\
&\geq&\left(\frac{M}{N}(a+1)-\frac{1}{2}\right)-\left(\frac{M}{N}a+\frac{1}{2}\right)-\frac{M}{N}+1\\
&=&0
\end{eqnarray}
giving that the open intervals are disjoint. Thus taking the
integers in the intervals $\bmod\;M$ remains disjoint (which
requires $i_1, i_2\leq N-1$).
\end{proof}

The second error which propagates throughout the proofs in
\Hales~and \Hallgren~stems from misconceptions about the division
operation $\Delta$. Both papers treat the image of the $\Delta$
map as a cartesian product, that is, the range on $t$ is the same
for all values $s$ ($M=32$ and $N=5$ illustrates how this fails to
give a bijection with their definitions). However, the image is
not a cartesian product; the values $t$ assumes depend on $s$,
otherwise we would have that $M$ is a multiple of $N$. In other
words, the cardinality of $C_s$ depends on $s$, with bounds given
in the following lemma, where we show that our definition works
and list some properties:

\begin{lemma}\label{l:deltaProp}
Using the notation from definition \emph{\ref{d:delta}},

1) the map $\Delta$ is well defined, and a bijection with its
image,

2) $\alpha=\beta+1$,

3) the sets of integers satisfy $\{-\beta,\dots,\beta\}\subseteq
C_s\subseteq\{-\alpha,\dots,\alpha\}$ for all
$s\in\{0,1,\dots,N-1\}$.
\end{lemma}
\begin{proof}
Given a $k$ in $\{0,1,\dots,M-1\}$, let $\Delta(k)=(s,t)$. Clearly
$0\leq s\leq N-1$. Set $\alpha=\floor{\frac{M}{2N}+\frac{1}{2}}$.
To check that $-\alpha\leq t\leq \alpha$, note
\begin{equation}
\frac{N}{M}k-\frac{1}{2}\leq k'\leq\frac{N}{M}k+\frac{1}{2}
\end{equation}
giving
\begin{equation}
\frac{M}{2N}+\frac{1}{2}\geq t=k-\round{\frac{M}{N}k'}\geq
-\left(\frac{M}{2N}+\frac{1}{2}\right)
\end{equation}
and $t$ integral allows the rounding operation. Thus the
definition makes sense.

Next we check that both forms of $\Delta$ in the definition are
bijections. Suppose $k_1\neq k_2$ are both in $\{0,1,\dots,M-1\}$,
with images $\Delta(k_r)=(s_r,t_r),r=1,2$. Let
$k_r'=\round{\frac{N}{M}k_r},r=1,2$. Note $0\leq k_r'\leq N$.

Assume $(s_1,t_1)=(s_2,t_2)$. If $k_1'=k_2'$, then
\begin{align}
t_1&=k_1-\round{\frac{M}{N}k_1'}=k_1-\round{\frac{M}{N}k_2'}\\
&\neq k_2-\round{\frac{M}{N}k_2'}= t_2
\end{align}
a contradiction. So we are left with the case $k_1'\neq k_2'$. In
order for $s_1=s_2$ we have (without loss of generality) $k_1'=0,
k_2'=N$. But then $t_1=k_1\geq 0$ and $t_2=k_2-M\leq M-1-M=-1$, a
contradiction. Thus $\Delta$ in the first sense is a bijection.

The second interpretation follows easily, since $-\alpha\leq
t\leq\alpha$ gives $0\leq t+\alpha\leq2\alpha$. So the second
register needs to have a basis with at least $2\alpha+1$ elements,
which causes the number of qubits needed\footnote{This is proven
in theorem \ref{t:MainBound}.} to implement the algorithm to be
$\ceil{\log M} + 2$ instead of $\ceil{\log M}$.

To see $\alpha=\beta+1$, bound $\alpha-\beta$ using the methods
above, and\footnote{$(M,N)=1$ is used to get the strict
inequalities.} one obtains $2>\alpha-\beta>0$.

All integers between $\round{\frac{M}{N}(s+1)}$ and
$\round{\frac{M}{N}s}$ inclusive must be of the form
$t_1+\round{\frac{M}{N}s}$ for $t_1\in C_s$ or of the form
$t_2+\round{\frac{M}{N}(s+1)}$ for $t_2\in C_{s+1}$. This range
contains $\round{\frac{M}{N}(s+1)} - \round{\frac{M}{N}s}+1\geq
\frac{M}{N}$ integers, and at most $\alpha+1$ of these are of the
form $t_2+\round{\frac{M}{N}(s+1)}$ with $t_2\in C_{s+1}$. This
leaves at least
$\ceil{\frac{M}{N}}-\alpha\geq\frac{M}{2N}-\frac{3}{2}$ that have
to be of the form $t_1+\round{\frac{M}{N}s}$ with $t_1\in C_s$,
implying $\beta\in C_s$. Similar arguments give $\pm \beta\in
C_s$, thus $\{-\beta,\dots,\beta\}\subseteq
C_s\subseteq\{-\alpha,\dots,\alpha\}$ for all $s$.
\end{proof}

$\Delta$ is efficient to implement as a quantum operation, since
it is efficient classically \cite[Chapter 4]{ChuangNielsen}.
Finally we note that $\Delta$, being a bijection, can be extended
to a permutation of basis vectors $\ket{k}$, thus can be
considered an efficiently implementable unitary operation.

We define some vectors we will need. For
$i\in\left\{0,1,\dots,N-1\right\}$ define
\begin{eqnarray*}
\ket{A^i} &=& F_M F_{LN}^{-1}\ket{Li}\\
&=& \frac{1}{\sqrt{LMN}}\sum_{k=0}^{M-1}\sum_{a=0}^{LN-1}\omega_{N}^{-ai}\omega_M^{ak}\ket{k}\\
\ket{B^i} &=& \ket{A^i}\text{ restricted to integers in the set } (i)\\
&=&\sum_{b\in(i)}A^i_b\ket{b}\\
&=& \frac{1}{\sqrt{LMN}}\sum_{b\in(i)}\sum_{a=0}^{LN-1}
\omega_{N}^{-ai}\omega_M^{ab}\ket{b}\\
\ket{T^i} &=& \ket{A^i}\text{ restricted to integers outside the set } (i)\\
&=&\sum_{b\not\in(i)}A^i_b\ket{b}\\
&=&\ket{A^i}-\ket{B^i}\\
&=& \frac{1}{\sqrt{LMN}}\sum_{b\not\in(i)}\sum_{a=0}^{LN-1}
\omega_{N}^{-ai}\omega_M^{ab}\ket{b}\\
\end{eqnarray*}
Think $A^i$ for actual values, $B^i$ for bump functions, and $T^i$
for tail functions. Note that the coefficients $B^i_b$ and $T^i_b$
are just $A^i_b$ for $b$ in the proper ranges.

We also define three equivalent shifted versions of $\ket{B^0}$.
Note that to make these definitions equivalent we require the sets
$(i)$ to have the same cardinality. Let
$\ket{S^i}=\sum_{b\in(0)}B^0_b\ket{b+i'} =
\sum_{b\in(0)}A^0_b\ket{b+i'}=\sum_{b\in(i)}A^0_{b-i'}\ket{b}$,
where each $b\pm i'$ expression is taken $\bmod\;M$. The
$\ket{S^i}$ have \emph{disjoint support}, which follows from lemma
\ref{l:setProp}, and will be important for proving theorem
\ref{t:oddCyclicQFT}.

\section{The Algorithm} The algorithm takes a
unit vector (quantum state) $\ket{u}$ on $\ceil{\log N}$
qubits\footnote{Recall logs are base 2.}, does a Fourier transform
$F_L$, $L$ a power of two, on another register containing
$\ket{0}$ with $\ceil{\log M}-\ceil{\log N}+2$ qubits, to
create\footnote{Note it may be more efficient to apply the
Hadamard operator $H$ to each qubit in $\ket{0}$.} a
superposition, and then reindexes the basis to create $L$
(normalized) copies of the coefficients of $\ket{u}$, resulting in
$\ket{u_L}$. Then another power of two Fourier transform $F_M$ is
applied. The division $\Delta$ results in a vector very close to
the desired output $F_N\ket{u}$ in the first register, with
garbage in the second register (with some slight entanglement).
The point of this paper is to show how close the output is to this
tensor product. We use $\ceil{\log M}+2$ qubits, viewed in two
ways: as a single register \ket{k}, or as a $\ceil{\log N}$ qubit
first register, with the remaining qubits in the second register,
written \ket{s}\ket{t}. We note that merely $\ceil{\log M}$ qubits
may not be enough qubits to hold some of the intermediate results.
The algorithm is:

\subsection{The odd cyclic quantum Fourier transform algorithm}\label{s:algorithm}
\begin{eqnarray}
\ket{u}\ket{0}&\xrightarrow{F_L}&\frac{1}{\sqrt{L}}
\sum_{i=0}^{N-1}\sum_{j=0}^{L-1}u_i\ket{i}\ket{j}\\
&\xrightarrow{\text{multiply}}&\frac{1}{\sqrt{L}}\sum_{i,j}u_i\ket{i+jN}\\
&=&\ket{u_L}\\
&\xrightarrow{F_M}&\frac{1}{\sqrt{LM}}\sum_{i,j}\sum_{k=0}^{M-1}u_i\omega_M^{\left(i+jN\right)k}\ket{k}\\
&\xrightarrow{\Delta}&\frac{1}{\sqrt{LM}}\sum_{i,j}u_i
\sum_{s=0}^{N-1}\sum_{t\in C_s}
\omega_M^{\left(i+jN\right)\left(t+\round{\frac{M}{N}s}\right)}\ket{s}\ket{t+\alpha}\\
&=&\frac{1}{\sqrt{N}}\sum_{i,s=0}^{N-1}u_i\omega_N^{is}\ket{s}\sqrt{\frac{N}{LM}}
\sum_{t\in C_s}\sum_{j=0}^{L-1}
\omega_M^{\left(i+jN\right)\left(t+\delta_s\right)}\ket{t+\alpha}\label{e:output}\\
&=&\ket{v}
\end{eqnarray}

$\ket{u_L}$ is the vector that is $L$ copies of the coefficients
from $\ket{u}$, normalized. $\ket{v}$ is the algorithm output.

Notice that $F_N\ket{u}$ appears in the output in line
\ref{e:output}, but the rest is unfortunately dependent on $s$ and
$i$. However the dependence is small: if $C_s$ were the same for
all $s$, if the $\delta_s$, which are bounded in magnitude by
$\frac{1}{2}$, were actually zero, and if the $i$ dependence were
dropped, then the output would leave $F_N\ket{u}$ in the first
register. The paper shows this is approximately true, and
quantifies the error.

\section{Initial bounds}We need many bounds to
reach the final theorem, which we now begin proving. \Hales~makes
the mistake of missing the $-1$ in the following
lemma\footnote{Using the definitions in \Hales, a $-\frac{1}{2}$
instead of -1 is sufficient. Even then, however, $M=128$, $N=37$,
$i=12$, and $k=40$ shows the error. Compare our lemma
\ref{l:Mbounds} to the proof of claim 4, section 9.2.3, in
\Hales.}; \Hallgren, using a different definition for the $(i)$,
is correct in dropping the $-1$. To avoid these subtle errors we
thus prove
\begin{lemma}\label{l:Mbounds}
For integers $N>2$, $M\geq 2N$, and any $i\in\{0,1,\dots,N-1\}$,
$k\in\{0,1,\dots,M-1\}$, with $k\not\in(i)$, we have
\begin{eqnarray}
\left|k-\frac{M}{N}i\right|_M&\geq&\frac{M}{2N}-1
\end{eqnarray}
\end{lemma}
\begin{proof}
The sets $(i)$ are disjoint, so we do two cases. If $i=0$, then
$k\not\in(0)$ implies
\begin{equation}
\frac{M}{2N}-\frac{1}{2}\leq k \leq M-\frac{M}{2N}+\frac{1}{2}
\end{equation}
from which it follows that
\begin{equation}
\left|k-\frac{M}{N}0\right|_M\geq\frac{M}{2N}-\frac{1}{2}>\frac{M}{2N}-1
\end{equation}

If $i\neq 0$, then either $k$ is less than the integers in $(i)$
or greater than the integers in $(i)$, giving two subcases.
Subcase 1:
\begin{equation}
0\leq k\leq\round{\frac{M}{N}i}-\frac{M}{2N}+\frac{1}{2}
\leq\frac{M}{N}i-\frac{M}{2N}+1
\end{equation}
implying
\begin{equation}
\frac{M}{2N}-1\leq\frac{M}{N}i-k\leq \frac{M}{N}i\leq
M-\frac{M}{N}
\end{equation}
which gives the bound. Subcase 2 is then
\begin{equation}
\frac{M}{N}i+\frac{M}{2N}-1\leq
\round{\frac{M}{N}i}+\frac{M}{2N}-\frac{1}{2}\leq k \leq M-1
\end{equation}
which implies
\begin{equation}
\frac{M}{2N}-1\leq k-\frac{M}{N}i \leq M-1-\frac{M}{N}i
\end{equation}
giving the bound and the proof.
\end{proof}

We now bound many of the $\ket{A^i}$ coefficients. Our bound has a
factor of $\pi$ not in \Hales~and \Hallgren, making it somewhat
tighter, and we avoid special cases\footnote{\Hales~and
\Hallgren~missed these cases by not placing any restriction such
as our hypothesis that $\frac{k}{M}-\frac{i}{N}$ is non-integral.
Compare our lemma \ref{l:approx} with Observation 2, section 9.2.3
in \Hales~and with Observation 1, section 3.1, in \Hallgren. }
where the statement would not be true.

\begin{lemma}\label{l:approx} For
$k\in\left\{0,1,\dots,M-1\right\}$ and
$i\in\left\{0,1,\dots,N-1\right\}$, with
$\frac{k}{M}-\frac{i}{N}$ not an integer, then
\begin{equation}
\left|A^i_k\right|\leq\sqrt{\frac{M}{LN}}\;\frac{2}{\pi\left|k
-\frac{M}{N}i\right|_M}
\end{equation}
\end{lemma}
\begin{proof}
We rewrite from the definition
\begin{eqnarray}
A^i_k &=& \frac{1}{\sqrt{LMN}}\sum_{a=0}^{LN-1}
\omega_M^{a\left(k-\frac{M}{N}i\right)}\\
\end{eqnarray}
which is a geometric series. By hypothesis,
$\omega_M^{\left(k-\frac{M}{N}i\right)}\neq 1$, so we can sum
as\footnote{Without this requirement, the sum would be $LN$, much
different than the claimed sum. The hypotheses avoid the resulting
divide by zero.}

\begin{eqnarray}
\left|A^i_k\right| &=&
\frac{1}{\sqrt{LMN}}\left|\frac{1-\omega_M^{LN\left(k-\frac{M}{N}i\right)}}
{1-\omega_M^{\left(k-\frac{M}{N}i\right)}}\right|
\end{eqnarray}
The numerator is bounded above by 2, and the denominator satisfies
\begin{eqnarray}
\left|1-\omega_M^{\left(k-\frac{M}{N}i\right)}\right| &=&
\left|1-\omega_M^{\left|k-\frac{M}{N}i\right|_M}\right|\\
&\geq& \frac{\pi \left|k-\frac{M}{N}i\right|_M}{M}
\end{eqnarray}
These together give
\begin{eqnarray}
\left|A^i_k\right| &\leq& \sqrt{\frac{M}{LN}}\;\frac{2}{\pi
\left|k-\frac{M}{N}i\right|_M}
\end{eqnarray}
\end{proof}

\TODO{add (M,N)=1 to some theorems}

Note our initial requirement that $(M,N)=1$ is strong enough to
satisfy the non-integral hypothesis in lemma \ref{l:approx},
except for the case $i=k=0$, which we will avoid.

Next we bound a sum of these terms. We fix
$\gamma=\frac{1}{2}-\frac{N}{M}$ for the rest of this paper.

\begin{lemma}\label{l:sumGeneral}Given integers $N>2$ and $M>2N$, with $N$ odd. Let
$\gamma=\frac{1}{2}-\frac{N}{M}$. For a fixed integer
$k\in\left\{0,1,\dots,M-1\right\}$,
\begin{equation}
\sum_{\substack{i=0\\k\not\in(i)}}^{N-1}\frac{1}{\left|k-\frac{M}{N}i\right|_M}
\leq\frac{2N}{M}\left(\frac{1}{\gamma}+\ln\left|\frac{N-1}{2\gamma}+1\right|\;\right)
\end{equation}
\end{lemma}
\begin{proof}The minimum value of the
denominator is at least $\frac{M}{2N}-1$ by lemma \ref{l:Mbounds},
and the rest are spaced out by $\frac{M}{N}$, but can occur
twice\footnote{Both \Hales~and \Hallgren~appear to overlook this
fact.} since the denominator is a sawtooth function going over one
period, giving that
\begin{eqnarray}
\sum_{\substack{i=0\\k\not\in(i)}}^{N-1}\frac{1}{\left|k-\frac{M}{N}i\right|_M}
&\leq&2\sum_{a=0}^{\frac{N-1}{2}}\frac{1}{\frac{M}{2N}-1+\frac{M}{N}a}\\
&=&\frac{2N}{M}\left(\frac{1}{\gamma}+\sum_{a=1}^{\frac{N-1}{2}}\frac{1}{\gamma+a}\right)\\
&\leq&\frac{2N}{M}\left(\frac{1}{\gamma}+\int_0^{(N-1)/2}\frac{1}{x+\gamma}dx\right)\\
&=&\frac{2N}{M}\left(\frac{1}{\gamma}+\ln\left|\frac{N-1}{2\gamma}+1\right|\;\right)
\end{eqnarray}
\end{proof}

The generality of the above lemma would be useful where physically
adding more qubits than necessary would be costly, since the lemma
lets the bound tighten as $\frac{N}{M}$ decreases. However the
following corollary is what we will use in the final theorem.

\begin{cor}\label{l:sum} Given integers $N\geq 13$ and $M\geq16N$, with $N$
odd.
For a fixed value $k\in\left\{0,1,\dots,M-1\right\}$,
\begin{equation}
\sum_{\substack{i=0\\k\not\in(i)}}^{N-1}\frac{1}{\left|k-\frac{M}{N}i\right|_M}
\leq\frac{4N\ln N}{M}
\end{equation}
\end{cor}
\begin{proof}
Using lemma \ref{l:sumGeneral}, $M\geq 16N$ gives
$\frac{1}{\gamma}\leq\frac{16}{7}$ and
\begin{eqnarray}
\frac{1}{\gamma}+\ln\left|\frac{N-1}{2\gamma}+1\right|&\leq&
\frac{16}{7}+\ln\left|\frac{8(N-1)}{7}+1\right|\\
&=&\ln\left(e^{\frac{16}{7}}\left(\frac{8(N-1)}{7}+1\right)\right)\\
&\leq& \ln\left(\frac{8}{7}\;e^{\frac{16}{7}}N\right)\\
&\leq&2\ln N
\end{eqnarray}
where the last step required $N\geq \left(
\frac{8}{7}\;e^{\frac{16}{7}}\right)>11.2$. The corollary follows.
\end{proof}

\Hales~claimed an incorrect bound\footnote{This fails, for
example, at $M=256$, $N=13$, $k=26$, using either their
definitions or our definitions.} of $\frac{2N\ln N}{M}$ in section
9.2.3, and \Hallgren~obtained the correct $\frac{4N\ln N}{M}$ in
section 3.1, but both made the errors listed above.

Next we prove a bound on a sum of the above terms, weighted with a
real unit vector. This will lead to a bound on the tails
$\left\|\sum_i\hat{u}_i\ket{T^i}\right\|$. Our bound has an extra
term compared to the claimed bounds in \Hales~and \Hallgren, but
corrects an error in their proofs.

\newcommand{\matbound}{\frac{22N\ln^2 N}{M}+\frac{32N^3}{M^2}}

\begin{lemma}\label{l:matNorm}
Given integers $N\geq 13$ and $M\geq16N$, with $N$ odd. For any
unit vector $x\in \mathbb{R}^N$
\begin{equation}
\sum_{k=0}^{M-1}\left|\sum_{\substack{i=0\\k\not\in(i)}}^{N-1}\frac{x_i}{\left|k-\frac{M}{N}i\right|_M}\right|^2\leq
\matbound\label{e:matBound}
\end{equation}
\end{lemma}

\begin{proof} We split the expression into three parts, the first
of which we can bound using methods from \Hales~and \Hallgren, and
the other two terms we bound separately.

Using the $\Delta$ operator from definition \ref{d:delta}, along
with the values $\alpha$ and $\beta$ defined there, and using
lemma \ref{l:deltaProp}, we can rewrite each $k$ with
$k=t+\round{\frac{M}{N}k'}=t+\frac{M}{N}k'+\delta_s$. Since $s$
differs from $k'$ by a multiple of $N$, and the $|x|_M$ function
has period $M$, in $\left|\frac{M}{N}(k'-i)+t+\delta_s\right|_M$
we can replace $k'$ with $s$. Rewrite the left hand side of
inequality \ref{e:matBound} as
\begin{align}
\sum_{k=0}^{M-1}\left|\sum_{\substack{i=0\\k\not\in(i)}}^{N-1}\frac{x_i}{\left|k-\frac{M}{N}i\right|_M}\right|^2&=
\sum_{s=0}^{N-1}\sum_{t\in C_s}\left|\sum_{\substack{i=0\\s\neq
i}}^{N-1}\frac{x_i}{\left|\frac{M}{N}(s-i)+t+\delta_s\right|_M}\right|^2
\end{align}
Letting $\Delta k=(s,t)$, note that $k\not\in(i)$ if and only if
$s\neq i$, which can be shown from the definitions and the
rounding rules used earlier. To simplify notation, write
$q_{i,s}^t=\frac{M}{N}(s-i)+t+\delta_s$. We have not changed the
values of the denominators, so $|q_{i,s}^t|_M\geq\frac{M}{2N}-1$
by lemma \ref{l:Mbounds} for all $i,(s,t)$ in this proof.

We want to swap the $s$ and $t$ sums, but we need to remove the
$t$ dependence on $s$. Again using lemma \ref{l:deltaProp}, we can
split the expression into the three terms\footnote{The second two
terms are missed in \Hales~and \Hallgren.}:
\begin{align}
\sum_{t=-\beta}^{\beta}\sum_{s=0}^{N-1}\left|\sum_{\substack{i=0\\s\neq
i}}^{N-1}\frac{x_i}{\left|q_{i,s}^t\right|_M}\right|^2\label{e:term1}\\
+\sum_{s\text{ with }\alpha\in
C_s}\left|\sum_{\substack{i=0\\s\neq
i}}^{N-1}\frac{x_i}{\left|q_{i,s}^\alpha\right|_M}\right|^2\label{e:term2}\\
+\sum_{s\text{ with }-\alpha\in
C_s}\left|\sum_{\substack{i=0\\s\neq
i}}^{N-1}\frac{x_i}{\left|q_{i,s}^{-\alpha}\right|_M}\right|^2\label{e:term3}
\end{align}
Next we bound the first term \ref{e:term1}. For a unit vector $x$
and fixed $t$ we rewrite the $s,i$ sum as the norm of a square
matrix $P_t$ acting on $x$, so that the sum over $s$ and $i$
becomes
\begin{align}
\left\|P_tx\right\|^2&=\sum_{s=0}^{N-1}\left|\sum_{\substack{i=0\\s\neq
i}}^{N-1}\frac{x_i}{\left|q_{i,s}^t\right|_M}\right|^2
\end{align}
We also define similarly to each $P_t$ a matrix $Q_t$ which is the
same except for minor modifications to the denominator:
\begin{align}
\left\|Q_tx\right\|^2&=\sum_{s=0}^{N-1}\left|\sum_{\substack{i=0\\s\neq
i}}^{N-1}\frac{x_i}{\left|q_{i,s}^t-\delta_s\right|_M}\right|^2
\end{align}
Note this matrix is circulant\footnote{That is, each row after the
first is the cyclic shift by one from the previous row.}, since
each entry in the matrix only depends on $s-i$. Also each entry is
nonnegative\footnote{$|q_{i,s}^t-\delta_s|_M\geq|q_{i,s}^t|_M-\frac{1}{2}\geq\frac{M}{2N}-\frac{3}{2}>0$
since $M>3N$}. Thus the expression is maximized by the vector
$y=\frac{1}{\sqrt{N}}\left(1,1,\dots,1\right)$ as shown in each of
\Hales, \Hallgren, and \cite{Hoy00}. Now we relate these matrix
expressions. Recall $|q_{i,s}^t|_M\geq\frac{M}{2N}-1$ and
$|\delta_s|\leq\frac{1}{2}$. Set $\lambda=\frac{N}{M-2N}$. Then we
find lower and upper bounds
\begin{align}
1-\lambda=1-\frac{1}{2(\frac{M}{2N}-1)}\leq
\frac{\left|q_{i,s}^t\right|_M-\frac{1}{2}}{\left|q_{i,s}^t\right|_M}
\leq\frac{\left|q_{i,s}^t-\delta_s\right|_M}{\left|q_{i,s}^t\right|_M}
\end{align}
and
\begin{align}
\frac{\left|q_{i,s}^t-\delta_s\right|_M}{\left|q_{i,s}^t\right|_M}
\leq\frac{\left|q_{i,s}^t\right|_M+\frac{1}{2}}{\left|q_{i,s}^t\right|_M}
\leq 1+\frac{1}{2(\frac{M}{2N}+1)} =1+\lambda
\end{align}
Rewriting
\begin{align}
\left\|P_tx\right\|^2&=\sum_{s=0}^{N-1}\left|\sum_{\substack{i=0\\s\neq
i}}^{N-1}\frac{x_i}{\left|q_{i,s}^t-\delta_s\right|_M}
\frac{\left|q_{i,s}^t-\delta_s\right|_M}
{\left|q_{i,s}^t\right|_M}\right|^2
\end{align}
and using the bounds gives
\begin{align}
(1-\lambda)^2\left\|Q_tx\right\|^2\leq\left\|P_tx\right\|^2\leq
(1+\lambda)^2\left\|Q_tx\right\|^2
\end{align}
Then since $y$ maximizes $\left\|Q_tx\right\|^2$,
\begin{align}
\left\|P_tx\right\|^2\leq
(1+\lambda)^2\left\|Q_tx\right\|^2\leq(1+\lambda)^2\left\|Q_ty\right\|^2
\leq\left(\frac{1+\lambda}{1-\lambda}\right)^2\left\|P_ty\right\|^2
\end{align}
giving that we can bound the leftmost term by
$\left(\frac{1+\lambda}{1-\lambda}\right)^2$ times the norm at
$y$. $\left(\frac{1+\lambda}{1-\lambda}\right)^2$ takes on values
between 1 and $\frac{225}{169}\approx 1.33$ for $M\geq 16N$,
better than the constant 4 in \Hales~ and \Hallgren.

Combined with corollary \ref{l:sum} this allows us to bound term
\ref{e:term1}:
\begin{align}
\sum_{t=-\beta}^{\beta}\sum_{s=0}^{N-1}\left|\sum_{\substack{i=0\\s\neq
i}}^{N-1}\frac{x_i}{\left|q_{i,s}^t\right|_M}\right|^2&\leq
\sum_t\frac{225}{169}\sum_{s=0}^{N-1}\left
|\sum_{\substack{i=0\\s\neq
i}}^{N-1}\frac{\frac{1}{\sqrt{N}}}{\left|q_{i,s}^t\right|_M}\right|^2\\
&\leq\left(2\beta+1\right)\frac{225}{169} \frac{N}{N} \left(
\frac{4N\ln N}{M}\right)^2\\
&\leq\frac{M}{N}\frac{225}{169} \left(
\frac{4N\ln N}{M}\right)^2\\
&\leq\frac{22N\ln^2 N}{M}
\end{align}

Now we bound the other two terms, \ref{e:term2} and \ref{e:term3}.
We need the following fact, which can be shown with calculus: the
expression $\left|\sum_{i=0}^{N-1}a_ix_i\right|$ subject to the
condition $\sum_{i=0}^{N-1}x_i^2=1$, has maximum value
$\sqrt{\sum_{i=0}^{N-1}a_i^2}$. Then term \ref{e:term2} can be
bounded using a similar technique as in the proof of lemma
\ref{l:sum}. Again we take $\gamma=\frac{1}{2}-\frac{N}{M}$.
\begin{align}
\sum_{s\text{ with }\alpha\in C_s}\left|\sum_{\substack{i=0\\s\neq
i}}^{N-1}\frac{x_i}{\left|q_{i,s}^\alpha\right|_M}\right|^2&\leq
\sum_{s}\left|\sqrt{\sum_{\substack{i=0\\s\neq
i}}^{N-1}\frac{1}{\left|q_{i,s}^\alpha\right|_M^2}}\right|^2\\
&\leq
N\frac{2N^2}{M^2}\left(\frac{1}{\gamma^2}+\sum_{a=1}^{\frac{N-1}{2}}\frac{1}
{\left(\frac{1}{2}-\frac{N}{M}+a\right)^2}\right)\\
&\leq \frac{2N^3}{M^2}\left(\frac{1}{\gamma^2}+\frac{1}{\gamma}
-\frac{1}{\frac{N-1}{2}+\gamma}\right)\\
&\leq\frac{16N^3}{M^2}
\end{align}
Term \ref{e:term3} is bound with the same method and result, and
adding these three bounds gives the desired inequality
\ref{e:matBound}.
\end{proof}

Similar to the proof of the previous lemma, Both \Hales~and
\Hallgren~claim the following bound
\begin{equation}
\sum_{k=0}^{M-1}\left|\sum_{\substack{i=0\\k\not\in(i)}}^{N-1}\frac{x_i}{\left|k-\frac{M}{N}i\right|_M}\right|^2\leq
\frac{4}{N}\sum_{k=0}^{M-1}\left|\sum_{\substack{i=0\\k\not\in(i)}}^{N-1}\frac{1}{\left|k-\frac{M}{N}i\right|_M}\right|^2
\end{equation}
leading to (in their papers)
\begin{align}
\sum_{k=0}^{M-1}\left|\sum_{\substack{i=0\\k\not\in(i)}}^{N-1}\frac{x_i}{\left|k-\frac{M}{N}i\right|_M}\right|^2&\leq
\frac{64N\ln^2N}{M}
\end{align}
So our bound tightens their 64 to a 22, but has a new term
accounting for the extra pieces in the proof. However, both
\Hales~(section 9.2.4) and \Hallgren~(appendix C) had the
following flaws their proofs. Both proofs rearranged the left
expression to be bounded by a matrix norm, and then ``rearranged"
the matrix to be square. This fails due to the subtle nature of
the $\Delta$ operation they implicitly used. They claimed the
resulting matrix differed only slightly from their previous one,
which is false, since many terms may have to be changed from 0 to
a large value. They relied on the resulting matrix being circulant
and being close to the to their initial expression, which it is
not due to these extra terms. Our proof above is based on their
methods, but avoids the errors they made by pulling out the
incorrect terms and bounding them separately, resulting in the
extra term in our bound.

We now use these lemmata to bound the tails
$\left\|\sum_i\hat{u}_i\ket{T^i}\right\|$. The bound $\frac{8\ln
N}{\sqrt{L}}$ was claimed in \Hales, section 9.2.3, and \Hallgren,
section 3.1, but our new terms from lemma \ref{l:matNorm} give us
a more complicated bound:

\newcommand{\tbound}{\frac{2}{\pi}\sqrt{\frac{22\ln^2
N}{L}+\frac{32N^2}{LM}}}

\begin{lemma}\label{l:tailBound}
Given three integers: an odd integer $N\geq 13$, $L\geq 2$ a power
of two, and $M\geq 16N$ a power of two, then
\begin{equation}
\left\|\sum_{i=0}^{N-1}\hat{u}_i\ket{T^i}\right\|\leq\tbound
\end{equation}
\end{lemma}
\begin{proof}
\begin{eqnarray}
\left\|\sum_{i=0}^{N-1}\hat{u}_i\ket{T^i}\right\|^2&=&
\sum_{k=0}^{M-1}\left|\sum_{\substack{i=0\\k\not\in(i)}}^{N-1}
\hat{u}_iT^i_k\right|^2\\
&\leq&\sum_k\frac{4M}{\pi^2LN}\left(\sum_{\substack{i=0\\k\not\in(i)}}^{N-1}
\frac{|\hat{u}_i|}{\left|k-\frac{M}{N}i\right|_M}\right)^2\label{e:bound1}\\
&\leq&\frac{4M}{\pi^2LN}\left(\matbound\right)
\end{eqnarray}
Taking square roots gives the result. Note that the requirements
of lemma \ref{l:approx} are satisfied when obtaining line
\ref{e:bound1}, since we avoid the $k=i=0$ case, and $(M,N)=1$.
\end{proof}

Next we show that the shifted $\ket{S^i}$ are close to the
$\ket{B^i}$, which will allow us to show the algorithm output is
close to a tensor product. This mirrors \Hales~claim 5, section
9.2.1, and \Hallgren~claim 2, section 3. In both cases their
constant was 4, where we obtain the better bound
$\frac{\pi}{\sqrt{3}}$.

\begin{lemma}\label{l:bumpApprox}
\begin{equation}
\Big\|\ket{S^i}-\ket{B^i}\Big\|\leq\frac{\pi LN}{M\sqrt{3}}
\end{equation}

\end{lemma}
\begin{proof}

Recall $\ket{S^i}=\sum_{b\in(i)}A^0_{b-i'\bmod M}\ket{b}$ and
$\ket{B^i}=\sum_{b\in(i)}A^i_b\ket{b}$. It is important these are
supported on the same indices! Also recall that
$\ket{A^i}=F_MF_{LN}^{-1}\ket{Li}$ and that $F_M$ is unitary. Then
(dropping $\bmod\;M$ throughout for brevity)
\begin{eqnarray}
\Big\|\ket{S^i}-\ket{B^i}\Big\|^2&=& \Big\|\sum_{b\in(i)}A^0_{b-i'}\ket{b}-\sum_{b\in(i)}A^i_b\ket{b}\Big\|^2\\
&\leq&\Big\|\sum_{k=0}^{M-1}A^0_{k-i'}\ket{k}-\sum_{k=0}^{M-1}A^i_k\ket{k}\Big\|^2\label{e:sameIndices}\\
&=&\Big\|F_M^{-1}\left(\sum_{k=0}^{M-1}A^0_k\ket{k+i'}-\ket{A^i}\right)\Big\|^2\\
&=&\sum_{a=0}^{LN-1}\left|\frac{1}{\sqrt{LN}}\omega_M^{-ai'}-\frac{1}{\sqrt{LN}}\omega_N^{-ai}\right|^2\\
&=&\frac{1}{LN}\sum_{a=0}^{LN-1}\left|\omega_M^{-ai'}\left(1-\omega_M^{a\delta_i}\right)\right|^2
\end{eqnarray}
and this can be bounded by
\begin{eqnarray}
\frac{1}{LN}\sum_{a=0}^{LN-1}\left|\frac{2\pi a
\delta_i}{M}\right|^2 \leq \frac{\pi^2}{LNM^2}\sum_{a=0}^{LN-1}a^2
\leq\frac{\pi^2}{LNM^2}\frac{(LN)^3}{3}
\end{eqnarray}
Taking square roots gives the bound.
\end{proof}

In the above proof, to obtain line \ref{e:sameIndices} we needed
that \ket{S^i} and \ket{B^i} have the same support, but \ket{S^i}
is a shifted version of \ket{B^0}, so we implicitly needed all the
sets $(i)$ to have the same cardinality. This is not satisfied in
\Hallgren~(although it is needed) but is met in \Hales.

\newcommand{\bdd}{\floor{\frac{M}{2N}-\frac{1}{2}}}
For the rest of he paper we need a set which is $(0)$ without
$\bmod\;M$ applied: let $\Lambda$ be those integers in the closed
interval $\left[-\bdd,\bdd\right]$. Then

\begin{lemma}\label{l:deltaKet}
\begin{eqnarray}
\Delta \ket{S^i}&=&\ket{i}\sum_{t\in \Lambda} A^0_t\ket{t+\alpha}
\end{eqnarray}
\end{lemma}
\begin{proof}
By definition,
$\ket{S^i}=\sum_{b\in(0)}A^0_b\ket{b+\round{\frac{M}{N}i} \bmod
M}$. $\Delta\left(b+\round{\frac{M}{N}i}\right)=(i,b)$ (the proof
uses $(M,N)=1$), and $\Delta$ a bijection implies
$\Delta\ket{b+\round{\frac{M}{N}i}\bmod M}=\ket{i}\ket{b+\alpha}$.
The rest follows\footnote{It is tempting to use $C_0$ instead of
$\Lambda$, but this is not correct in all cases.}.
\end{proof}

\section{Main results}\label{s:MainResults}
Now we are ready to use the above lemmata to prove the main
theorem.

\begin{theorem}\label{t:oddCyclicQFT}
Given three integers: an odd integer $N\geq 13$, $L\geq 16$ a
power of two, and $M\geq LN$ a power of two. Then the output
$\ket{v}$ of the algorithm in section \emph{\ref{s:algorithm}}
satisfies
\begin{equation}
\Big\|\ket{v}-F_N\ket{u}\otimes\sum_{t\in
\Lambda}A^0_t\ket{t+\alpha}\Big\|\leq \tbound+\frac{\pi
LN}{M\sqrt{3}}
\end{equation}
\end{theorem}

\begin{proof}
Note
\begin{align}
\ket{\hat{u}}&:=F_N\ket{u}=\sum_{i=0}^{N-1}\hat{u}_i\ket{i}&
F_M\ket{u_L}&=\sum_{i=0}^{N-1}\hat{u_i} \ket{A^i}
\end{align}

Using lemma \ref{l:deltaKet} and that $\Delta$ is unitary allows
us to rewrite the left hand side as
\begin{eqnarray}
\Big\|\ket{v}-\sum_{\substack{s=0\\t\in C_s}}^{N-1}\hat{u}_s
A^0_t\ket{s}\ket{t+\alpha}\Big\|
&=&\Big\|\Delta F_M\ket{u_L}-\sum_{s=0}^{N-1}\hat{u_s}\Delta\ket{S^s}\Big\|\\
&=&\Big\|\sum_{s=0}^{N-1}\hat{u_s}\ket{A^s}-\sum_{s=0}^{N-1}\hat{u_s}\ket{S^s}\Big\|\\
&=&\Big\|\sum_{s=0}^{N-1}\hat{u_s}(\ket{B^s}+\ket{T^s})-\sum_{s=0}^{N-1}\hat{u_s}\ket{S^s}\Big\|
\end{eqnarray}
By the triangle inequality this is bounded by
\begin{eqnarray}
\Big\|\sum_{s=0}^{N-1}\hat{u_s}\ket{T^s})\Big\|+
\Big\|\sum_{s=0}^{N-1}\hat{u_s}\ket{B^s})-\sum_{s=0}^{N-1}\hat{u_s}\ket{S^s}\Big\|
\end{eqnarray}
which in turn by lemmata \ref{l:tailBound} and \ref{l:bumpApprox}
is bounded by
\begin{eqnarray}
\tbound+\frac{\pi
LN}{M\sqrt{3}}\sqrt{\sum_s|\hat{u_s}|^2}\label{e:disjoint}
\end{eqnarray}
The last expression has $\left\|\ket{\hat{u}}\right\|=1$, which
gives the result. Note that to obtain line \ref{e:disjoint} we
needed the supports of the $\ket{B^s}$ disjoint, and that the
\ket{S^i} and \ket{B^i} have the same support\footnote{This is not
satisfied in \Hales, and the overlapping portions make that proof
invalid.}.
\end{proof}

This shows that the output of the algorithm in section
\ref{s:algorithm} is close to a tensor product of the desired
output $F_N\ket{u}$ and another vector (which is not in general a
unit vector). Since a quantum state is a unit vector, we compare
the output to a unit vector in the direction of our approximation
via:

\begin{lemma}\label{l:unitTriangle}
Let $\vec{a}$ be a unit vector in a finite dimensional vector
space, and $\vec{b}$ any vector in that space. For any
$0\leq\epsilon\leq 1$, if
$\left\|\vec{a}-\vec{b}\right\|\leq\epsilon$ then the unit vector
$\vec{b'}$ in the direction of $\vec{b}$ satisfies
$\left\|\vec{a}-\vec{b'}\;\right\|\leq\epsilon\sqrt{2}$.
\end{lemma}
\begin{proof}
Simple geometry shows the distance is bounded by
$\sqrt{2(1-\sqrt{1-\epsilon^2})}$, and this expression divided by
$\epsilon$ has maximum value $\sqrt{2}$ on $(0,1]$. The
$\epsilon=0$ case is direct.
\end{proof}

So we only need a $\sqrt{2}$ factor to compare the algorithm
output with a unit vector which is $F_N\ket{u}$ tensor another
unit vector. We let $\ket{\psi}$ denote the unit length vector in
the direction of $\sum_{t\in \Lambda}A^0_t\ket{t+\alpha}$ for the
rest of this paper\footnote{The subscripts in $A^0_t$ are taken
$\bmod M$.}.

For completeness, we repeat arguments from \cite{HalHal00,Hoy00}
to obtain the operation complexity and probability distribution,
and we show concrete choices for $M$ and $L$ achieving a desired
error bound.

To show that measuring the first register gives measurement
statistics which are very close to the desired distribution, we
need some notation. Given two probability distributions
$\mathcal{D}$ and $\mathcal{D}'$ over $\{0,1,\dots,M-1\}$, let
$\left|\mathcal{D}-\mathcal{D}'\right|=\sum_{k=0}^{M-1}\left|\mathcal{D}(k)-\mathcal{D}'(k)\right|$
denote the total variation distance. Then a result\footnote{Their
statement is a bound of $4\epsilon$, but their proof gives the
stronger result listed above. We choose the stronger form to help
minimize the number of qubits needed for simulations.} of
Bernstein and Vazirani \cite{BV97} states that if the distance
between any two states is small, then so are the
induced\footnote{The induced distribution from a state \ket{\phi}
is $\mathcal{D}(k)=|\braket{k}{\phi}|^2$.} probability
distributions:
\begin{lemma}[\cite{BV97}, Lemma 3.6]\label{l:prob}
Let \ket{\alpha} and \ket{\beta} be two normalized states,
inducing probability distributions $\mathcal{D}_\alpha$ and
$\mathcal{D}_\beta$. Then for any $\epsilon>0$
\begin{equation}
\left\|\ket{\alpha}-\ket{\beta}\right\|\leq\epsilon \Rightarrow
\left|\mathcal{D}_\alpha-\mathcal{D}_\beta\right|\leq
2\epsilon+\epsilon^2
\end{equation}
independent of what basis is used for measurement.
\end{lemma}

Combining this with theorem \ref{t:oddCyclicQFT} and lemmata
\ref{l:unitTriangle} and \ref{l:prob} gives the final result

\newcommand{\minA}{\ensuremath{65}}
\newcommand{\minB}{\ensuremath{735}}

\begin{thm}\label{t:MainBound}~\\

\emph{\textbf{1)}} Given an odd integer $N\geq 13$, and any
$\sqrt{2}\geq\epsilon>0$. Choose $L\geq 16$ and $M\geq L N$ both
integral powers of $2$ satisfying
\begin{equation}
\tbound+\frac{\pi LN}{M\sqrt{3}}\leq
\frac{\epsilon}{\sqrt{2}}\label{e:mainBound}
\end{equation}
Then there is a unit vector \ket{\psi} such that the output
\ket{v} of the algorithm in section \emph{\ref{s:algorithm}}
satisfies
\begin{equation}\label{e:errorBound}
\\|\ket{v}-F_N\ket{u}\otimes\ket{\psi}\|\leq\epsilon
\end{equation}

\emph{\textbf{2)}} We can always find such an $L$ and $M$ by
choosing
\begin{align}
L&=c_1\frac{\sqrt{N}}{\epsilon^2}\label{e:Lvalue}\\
M&=c_2\frac{N^\frac{3}{2}}{\epsilon^3}\label{e:Mvalue}
\end{align}
for some constants $c_1,c_2$ satisfying
\begin{align}
\minA \leq c_1 &\leq 2\times\minA\label{e:const1}\\
\minB \leq c_2 &\leq 2\times\minB\label{e:const2}
\end{align}

\newcommand{\mr}{\frac{\sqrt{N}}{\epsilon}}

\emph{\textbf{3)}} The algorithm requires $\ceil{\log M}+2$
qubits. By claim $2$ a sufficient number of qubits is then
$\ceil{12.53+3\log\frac{\sqrt{N}}{\epsilon}}$. The algorithm has
operation complexity $O(\log M(\log \log M + \log1/\epsilon))$.
Again using claim $2$ yields an operation complexity of
\begin{equation}
O\left(\log \mr\left(\log \log \mr + \log1/\epsilon\right)\right)
\end{equation}

\emph{\textbf{4)}} The induced probability distributions
$\mathcal{D}_v$ from the output and $\mathcal{D}$ from
$F_N\ket{u}\otimes\ket{\psi}$ satisfy
\begin{equation}
\left|\mathcal{D}_v-\mathcal{D}\right|\leq 2\epsilon+\epsilon^2
\end{equation}
\end{thm}

\begin{proof}
Claim 1 follows directly from theorem \ref{t:oddCyclicQFT} and
lemma \ref{l:unitTriangle}. Claim 1 and lemma \ref{l:prob} give
claim 4.

To get claim 2, note that for the bound to be met, we must have
$\frac{\ln^2 N}{L}<\epsilon^2$, $\frac{N^2}{LM}<\epsilon^2$, and
$\frac{LN}{M}<\epsilon$. Trying to keep $M$ small as $N$ and
$\epsilon$ vary leads to the forms for $L$ and $M$ chosen. If we
substitute lines \ref{e:Lvalue} and \ref{e:Mvalue} into
\ref{e:mainBound} and simplify, we get
\begin{align}
\frac{4}{\pi}\sqrt{\frac{11\ln^2N}{c_1\sqrt{N}}+
\frac{16\epsilon^3}{c_1c_2}}
+\frac{\pi\sqrt{2}}{\sqrt{3}}\frac{c_1}{c_2}&\leq 1
\end{align}
The left hand side is largest when $\epsilon=\sqrt{2}$ and $N=55$,
so it is enough to find constants $c_1$ and $c_2$ such that
\begin{align}
\frac{4}{\pi}\sqrt{\frac{11\ln^2 55}{c_1\sqrt{55}}+
\frac{32\sqrt{2}}{c_1c_2}}
+\frac{\pi\sqrt{2}}{\sqrt{3}}\frac{c_1}{c_2}&\leq
1\label{e:c1c2bound}
\end{align}
Ultimately we want $L$ and $M$ to be powers of two, so we find a
range for each of $c_1$ and $c_2$ such that the upper bound is at
least twice the lower bound, and such that all pairs of values
$(c_1,c_2)$ in these ranges satisfy inequality \ref{e:c1c2bound}.
To check that the claimed ranges work, note that for a fixed
$c_1$, the expression increases as $c_2$ decreases, so it is
enough to check the bound for $c_2=\minB$. After replacing $c_2$
in the expression with $\minB$, the resulting expression has first
and second derivatives with respect to $c_1$ over the claimed
range, and the second derivative is positive, giving that the
maximum value is assumed at an endpoint. So we only need to check
inequality \ref{e:c1c2bound} at two points:
$(c_1,c_2)=(\minA,\minB)$ and $(2\times\minA,\minB)$, both of
which work. Thus the bound is met for all $(c_1,c_2)$ in the
ranges claimed. With these choices for $M$ and $L$, note that
$L\geq 16$ and $M\geq LN\Leftrightarrow c_2\geq\epsilon c_1$,
which is met over the claimed range, so all the hypothesis for
claim 1 are satisfied.

Finally, to prove claim 3, algorithm \ref{s:algorithm} and the
proof of lemma \ref{l:deltaProp} give that we need $\ceil{\log N}$
qubits in the first register and $\max\{\ceil{\log
L},\ceil{\log(2\alpha+1)}\}$ qubits in the second register. $L\leq
\frac{M}{N}< 2\alpha+1$ gives that it is enough to have
$\ceil{\log(2\alpha+1)}$ qubits in the second register. Then
$2\alpha+1\leq \frac{M}{2N}+2$ gives
\begin{align}
\ceil{\log(2\alpha+1)}\leq\ceil{1+\log M - \log N} =2+\ceil{\log
M}-\ceil{\log N}
\end{align}
Thus $\ceil{\log M}+2$ is enough qubits\footnote{An example
requiring $\ceil{\log M}+2$ qubits is $M=1024$, $N=65$, so the
bound is tight.} for the algorithm. By claim 2, we can take $M\leq
2\times \minB \frac{N^{3/2}}{\epsilon^3}$ giving $\ceil{\log
M}+2\leq\ceil{12.53+3\log\frac{\sqrt{N}}{\epsilon}}$.

As noted in \Hales~and \Hallgren, the most time consuming step in
algorithm \ref{s:algorithm} is the $F_M$ Fourier computation.
Coppersmith \cite{Copper94} shows how to $\epsilon$ approximate
the quantum Fourier transform\footnote{Many authors give a simple
quantum circuit doing the quantum Fourier transform over a power
$2^m$ with time complexity $O(m^2)$; see for example \cite[Chapter
5 and endnotes]{ChuangNielsen}. However, this requires $m$
elementary operations, which seems a little like cheating.
Requiring a finite fixed number of elementary operations would
give a time complexity of $O(m^3)$.}
 for order $M=2^m$ with operation complexity of
$O(\log M(\log \log M + \log1/\epsilon))$. Using this to
approximate our approximation within error $\epsilon$ gives the
time complexities in claim 3, finishing the proof.
\end{proof}

\section{Conclusion}
These bounds allow simulation for many choices of $N$ and
$\epsilon$. However the choices for $M$ and $L$ given in theorem
\ref{t:MainBound} can usually be improved, and were merely given
to show such values can be found. For example, the following table
shows, for different $N$ and $\epsilon$ combinations, a triple
$(g,m,l)$ of integers, with the choice from line \ref{e:Mvalue}
being $M=2^g$; yet in each case $M=2^m$ and $L=2^l$ is the pair
with minimal $m$ satisfying the hypotheses for theorem
\ref{t:MainBound}. Thus choosing $M$ and $L$ carefully may allow
lower qubit counts, such as the $N=13$, $\epsilon=0.10$ case.
\begin{table}[h]
\begin{center}
\begin{tabular}{| r || r | r | r | r | r | r |}\hline

$\epsilon$ & N=13 & N=25 & N=51 & N=101 & N=251 & N=501
\\\hline\hline

.001&45,45,28&47,47,28&48,48,29&50,50,29&52,52,30&53,53,30
\\\hline

.01&36,35,21&37,37,22&38,38,23&40,40,23&42,42,23&43,43,24
\\\hline

.05&29,28,17&30,30,17&31,31,18&33,33,18&35,35,19&36,36,19
\\\hline

.10&26,25,15&27,27,15&28,28,16&30,30,16&32,32,17&33,33,17
\\\hline

.20&23,22,13&24,24,13&25,25,14&27,27,14&29,29,15&30,30,15
\\\hline

.30&21,20,12&22,22,12&24,24,12&25,25,13&27,27,13&29,28,14
\\\hline

.40&20,19,11&21,21,11&22,22,12&24,24,12&26,26,13&27,27,13
\\\hline

\end{tabular}\\
\caption{Values}\label{ta:values}
\end{center}
\end{table}

We also simulated this algorithm for the combinations above
requiring 22 or fewer qubits. The first test computed the
algorithm error on random input vectors (states)\footnote{The left
hand side of line \ref{e:errorBound} is the error computed.}. The
middle set of columns in table \ref{ta:test1}, where
$(M,L)=(2^m,2^l)$, shows the maximal error observed in the column
labelled ``observed $\epsilon$'' over 100 random vectors. Note
that the observed error is \emph{much} smaller than the required
bound; for example, with $N=25$, $\epsilon=0.3$ the max observed
error is actually 0.0182. This led to the second set of
experiments, with results in the third set of columns in table
\ref{ta:test1}, where we tried all legal $M,L$ combinations until
we found the one with smallest $M$ value that met the desired
error bound, when tested over 5000 random vectors. This seemed to
show that the qubit cost could almost be cut in half in practice.
As a final test, for $N=501$ and $\epsilon=0.2$, the theorem
requires 30 qubits, but empirical testing showed 15 suffices for
our 5000 test vectors, which had a maximal error of 0.18. These
results show that it is likely that significant tightening of the
bounds presented here is possible, resulting in qubit savings.

\begin{table}[h]
\begin{center}
\begin{tabular}{| c | c || c | c | r|| c | c | c ||}\hline
$N$ & $\epsilon$ & $m$ & $l$ & observed $\epsilon$ & {best $m$} &
best $l$ & $\epsilon_2$
\\\hline\hline

13 & 0.4 & 19 & 11 & 0.0362329 & 9 & 4 & 0.353615\\

13 & 0.3 & 20 & 12 & 0.0409662 & 10 & 4 & 0.212023\\

13 & 0.2 & 22 & 13 & 0.0187127 & 11 & 4 & 0.158535
\\\hline

25 & 0.4 & 21 & 11 & 0.0193478 & 10 & 4 & 0.309438\\

25 & 0.3 & 22 & 12 & 0.0181997 & 11 & 4 & 0.193214
\\\hline

51 & 0.4 & 22 & 12 & 0.0332493 & 11 & 4 & 0.294778
\\\hline

\end{tabular}\\
\caption{Simulation results}\label{ta:test1}
\end{center}
\end{table}

\bibliographystyle{amsplain}
\bibliography{qbib}


\end{document}